\documentclass[11pt]{article}
\usepackage[dvips]{graphicx}
\newcommand{\be}{\begin{eqnarray}}
\newcommand{\ee}{\end{eqnarray}}

\newcommand{\nd}{\noindent}

\begin{document} 
\begin{center}
{\bf \large Freeze-out volume of hot dense fireball}
\vskip 0.2in
{M. Mishra\footnote{Email: madhukar.12@mail.com} and C. P. Singh\footnote{Email: cpsingh\_bhu@yahoo.co.in}}\\
{\it Department of Physics, Banaras Hindu University, Varanasi - 221005, India}
\end{center}
\begin{center}
Abstract
\end{center}
     A thermodynamically consistent excluded volume model is proposed to account for the particle multiplicities obtained from lowest SIS energies to the highest RHIC energies. The chemical freeze-out volumes lying in a slice of one unit of rapidity for pions and kaons are separately inferred from this analysis and the results are compared with the corresponding thermal freeze-out volumes obtained from the Hanbury-Brown Twiss (HBT) pion interferometry. Furthermore, we extract the variations of freeze-out number densities for pions and nucleons with the center-of-mass energy in our model and compare them with the HBT data.   
\vskip 0.2in
{\nd \bf PACS numbers}: 25.75.-q; 25.75.Gz

{\nd \bf Keywords}: Excluded volume model, Equation of state, thermodynamical consistency, freeze-out volume.
\newpage
\section{INTRODUCTION}
   In the last few years experiments at Brookhaven and CERN have revealed information over a wide spectrum of observables obtained with Au+Au and Pb+Pb heavy ion colliders, respectively. These observations show that instead of behaving like a gas of free quarks and gluons, as was expected, the matter created in RHIC's heavy-ion collisions appears to be more like a perfect liquid [1]. This exists as a hot and dense fireball which subsequently expands, cools and finally disintegrates into various hadrons, leptons and photons etc. [2, 3]. Here one gets a surprising result that a consistent and suitable description of all the experimental results on particle multiplicities and particle ratios from the lowest SIS to the highest RHIC energies is available within the framework of a thermal statistical model [4, 5]. The situation becomes a little confusing because thermal models provide a description for the hot, dense hadron gas (HG) which may have resulted after QGP or without a QGP formation. Chemical equilibrium in HG removes any memory present regarding a phase transition. The purpose of this letter is to point out certain anomalous features which if present in the experimental data, can yield information regarding occurrence of a phase transition at RHIC energy.

     Although emissions of hadrons in a statistical model essentially involve the idea of equilibration in the hadron gas picture, it does not reveal any information regarding the existence of a QGP phase before hadronization. However, we know that a system which has gone through a mixed phase in a first order phase transition, is expected to possess a much larger space-time extent than what we expect from a system if it remains in the hadronic phase only. Thus the freeze-out volume $V_f$ of the fireball is a significant quantity which can give a hint whether a phase transition has occurred or not. It should be mentioned that so far this aspect of investigation has not been pursued seriously in the statistical (thermal) model calculations. An extremely large fireball volume at freeze-out is clearly indicative of the fact that an equilibrated phase of QGP existed before a chemically equilibrated hadron gas evolved during the ultra-relativistic heavy ion collisions at RHIC. A dense and hot hadron gas fireball does not retain the memory of the QGP phase existing before hadronization  if the resulting hadron gas achieves chemical equilibrium. However, an extremely large $V_f$ of the fireball can certainly indicate the existence of a long-lived mixed phase in a first order phase transition. A better understanding of the hadronic freeze-out is urgently required in order to fully comprehend the properties of the fireball, which is governed by parameters like freeze-out temperature $T_f$, baryon chemical potential $\mu_B$ and the fireball volume $V_f$. The point of chemical freeze-out fixes the chemical composition of the fireball and the inelastic collisions among the particles are stopped. After this stage the particles continue to interact through the elastic collisions until their momentum distribution is fixed and this point is called thermal freeze-out in the heavy-ion collision. The freeze-out volume throws light on the collective expansion as well as anisotropic flow of the fluid existing in the fireball before final thermal freeze-out. Recently Admova et al., have determined a universal condition [6] for thermal freeze-out of the pions in heavy-ion collisions by proposing that the mean-free path of pions at freeze-out is $\lambda_{\pi}=1$ fm and thus space-time extent of the source of pions at thermal freeze-out was determined from the pion interferometry (HBT) data. The data reveal the existence of a minimum in the variation of the freeze-out volume $V_f$  with $\sqrt{s_{NN}}$ between AGS and SPS energies and this feature poses a problem, which cannot be understood in terms of any freeze-out mechanism. The purpose of this note is to calculate the variation of $V_f$ with the center-of-mass energy in a thermal statistical model incorporating excluded volume for baryons and compare the predictions with that deduced from HBT data in order to have independent confirmation regarding correctness of the geometrical assumptions involved in these models on the space-time dynamics of the fireball. 

    HBT measurements reveal thermal (kinetic) freeze-out volume. When it is extrapolated to full phase space volume, this yields the upper limit for the fireball volume and hence we can infer about the space-time properties of the highly excited system created in the ultra-relativistic heavy-ion collisions. It was theoretically predicted that the presence of a long-lived mixed phase during a first order phase transition from a QGP to a gas of hadrons will result in a large source size and it can be measured by HBT interferometry [7-9]. However, if a system undergoes a collective expansion, the extracted HBT radii do not indicate the exact geometrical size of the pion source at freeze-out. Depending on the expansion rate, the HBT radii result from a combination of geometrical and thermal length scales (as given by the thermal velocity of the pions). The lack of an energy dependence of HBT radii and the inconsistency with the hydrodynamical models are collectively called as 'HBT puzzle' at RHIC [10]. Due to expansion, the HBT volume will be far less than the total volume of the system. This is a well known effect of expansion and, therefore, HBT volume can serve as a lower estimate of the total volume. The particle ratios as given in a thermal statistical model yield the chemical freeze-out parameters. If there is a significant time difference between the times of the kinetic and the chemical freeze-out, then the HBT volume might change during this period. We, therefore, want to emphasize that the comparison of two freeze-out volumes should be taken with care since there are many factors which when implemented, will change the results of HBT volume. Similarly determination of a chemical freeze-out volume from a thermal model calculation depends crucially on the excluded volume effect incorporated in the thermal HG model. 

     In this letter we propose a new excluded volume model which has many advantages over other models existing in the literature. Thus a comparison of the predictions of the model with HBT data will reveal many crucial information regarding the geometrical space-time extent of the fireball and the thermal parameters, e.g., $T_f,\; \mu_B$ and the energy density etc. and this will resolve many unsettled issues involved in heavy-ion physics.

\section{EXCLUDED VOLUME MODEL}
   Since ideal gas description is unsuitable for dealing with HG picture at very large temperature/density, usually excluded volume models have been used in describing a hot and dense HG [11-19]. We derive here a new thermodynamically consistent excluded volume model for the description of hot and dense hadron gas (HG). This model has been obtained from our old model [11] in which we consider baryons having an eigen volume $V_i$, also $R=\sum_i\,n_i^{ex}\,V_i$ is the fraction of occupied volume and $n_i^{ex}$ is the baryon density for $i^{th}$ species in the excluded volume approach. Thus we can write [11]:
\begin{equation}
n_i^{ex}=(1-R)\,I_i\,\lambda_i-I_i\,\lambda_i^2\,\frac{\partial R}{\partial\lambda_i}.
\end{equation}

Here $\lambda_i$ is the fugacity of the $i^{th}$ baryonic species and $I_i$  is the following integral in the partition function using Boltzmann approximation:
\begin{equation}
I_i=\frac{g_i}{6\pi^2\,T}\,\int_0^{\infty}\,\frac{k^4\,dk}{\sqrt{k^2+m_i^2}}\,exp(-\sqrt{k^2+m_i^2}/T).
\end{equation}

Here $n_i^0=I_i\,\lambda_i$ is the   baryon density if baryons are point-like particles. It is obvious that the presence of the $\partial R/\partial\lambda_i$ in Eq. (1) makes the expression thermodynamically consistent. If we put $\partial R/\partial\lambda_i=0$ and we take only one type of baryons, then $R=n_i^{ex}\,V_i$ and from Eq. (1) we get the thermodynamically inconsistent expression [12]:
\begin{equation}
n_i^{ex}=\frac{n_i^0}{1+n_i^0\,V_i}.
\end{equation}

Therefore, we can treat $\partial R/\partial\lambda_i$ as a small correction term and Eq. (1) can be rewritten in the form [11]:
\begin{equation}
R=(1-R)\,\sum_i\,n_i^0\,V_i-\sum_i\,n_i^0\,V_i\,\lambda_i\,\frac{\partial R}{\partial\lambda_i}.
\end{equation}

Taking $R^0=\sum_i\,X_i$, where $X_i=I_i\,V_i\,\lambda_i$ which means $R^0=\sum_i\,n_i^0\,V_i$ and putting $\partial R/\partial\lambda_i=0$, we get
\begin{equation}
R=\hat R=\frac{R^0}{1+R^0}.
\end{equation}
Thus we can write Eq. (4) in the form 
\begin{equation}
R=\hat R+\Omega R,
\end{equation}
where
\begin{equation}
\Omega=-\frac{1}{1+R^0}\,\sum_i\,I_i\,\lambda_i^2\,V_i\frac{\partial}{\partial\lambda_i}.
\end{equation}           
By using Neumann iteration method, the Eq. (6) can be written as:
\begin{equation}
R=\hat R+\Omega\hat R+\Omega^2\hat R+\Omega^3\hat R+\cdots
\end{equation}

By retaining terms up to $\Omega^2\hat R$ only, the expression for $R$ can be written as: 
\begin{equation}
R=\frac{\sum_i\,X_i}{1+\sum\,X_i}-\frac{\sum_i\,X_i^2}{(1+\sum_i\,X_i)^3}+
2\frac{\sum_i\,X_i^3}{(1+\sum_i\,X_i)^4}-3\frac{\sum_i\,X_i\,\lambda_i\sum_i\,
X_i^2\,I_i\,V_i}{(1+\sum_i\,X_i)^5}.
\end{equation}
               
Finally by calculating the values of $R$ and $\partial R/\partial\lambda_i$, one can calculate the value of particle number density by using Eq. (1).  Similarly the baryonic pressure can be given as:
\begin{equation}
P^{ex}=(1-R)\,\sum_i\,P_i^0.
\end{equation}
                                                                    
   Obviously this approach looks more simple and attractive in comparison to other excluded volume approaches which are thermodynamically consistent [11,13]. Moreover, this approach has an added advantage since it can be used for extremely low as well as extremely large values of temperature $T$ and baryon chemical potential $\mu_B$ where all the other approaches fail to give satisfactory results. It is noteworthy that in comparison to other models, our earlier model does not show violation of causality even in extreme cases of $T$ and $\mu_B$ [14]. 
   
   In order to parameterize center-of-mass energy $\sqrt{s_{NN}}$ in terms of $T$ and $\mu_B$, we use the following parameterization for $\mu_B$ and $T$  
\begin{eqnarray}
\mu_B=\frac{a}{1+b\,\sqrt{s_{NN}}},\\
T=c-d\,e^{-f\,\sqrt{s_{NN}}},
\end{eqnarray}
where $a=1.308\pm0.028$ GeV, $b=0.273\pm0.008$ GeV$^{-1}$ and $c=172.3\pm 2.8$ MeV, $d=149.5\pm 5.7$ MeV $f=0.20\pm0.03$ GeV [22-26]. We have also used one additional parameter as the hard-core volume $V$. The hard-core volume of all kinds of the baryons is taken as $V=0.8$ fm$^3$.

    We have considered all particles and resonances up to mass of $2$ GeV/c$^2$ in our calculation. Here resonances having well defined masses and widths have been incorporated. Branching ratios [27] for sequential decays have been suitably accounted and in the presence of several decay channels, only dominant channel has been used for the calculation. It should be mentioned that success of any thermal statistical model relies strongly on the input from the particle data table [27]. However, any change in the mass cut-off does not produce any appreciable change on the features of the curves obtained from this calculation. We have thus calculated the particle densities $\pi^{+},\;K^{+},\;K^{-}$ at several energies and after comparing with the experimental results of total multiplicities for these particles we have deduced the total chemical freeze-out volume of the fireball. We intend to give details of the comparisons with the experimental results regarding particle multiplicities and particle ratios in a subsequent publication. Here we wish to report that the curves of $\pi^{+},\;K^{+},\;K^{-}$       multiplicities are very well reproduced by our model.  
       
\begin{figure}[tbp]
\begin{center}
{\includegraphics[scale=0.8]{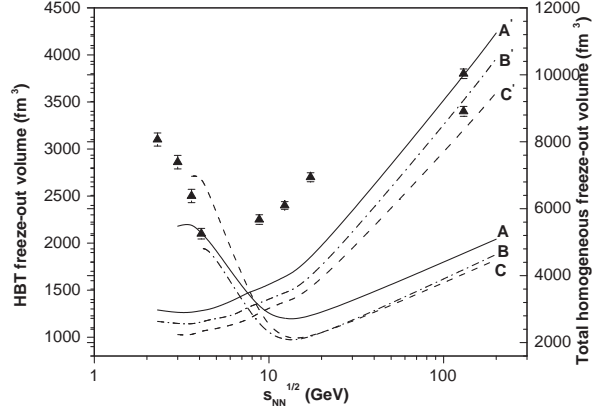}}
\caption{The freeze-out volume is represented by $A^{'},\, B^{'},\, C^{'}$ for our model for $\pi^{+},\, K^{+},\, K^{-}$, respectively as a function of center-of-mass energy $\sqrt{s_{NN}}$. We have also shown the volume in a slice of unit rapidity i.e., $dV/dy$ as calculated from our model  as shown by $A,\,B,\,C$ for $\pi^{+},\, K^{+},\, K^{-}$, respectively. HBT data for freeze-out volume $V_{HBT}$ for the $\pi^{+}$ are shown by the triangular points for comparison.}
\end{center}
\end{figure}

\begin{figure}[tbp]
\begin{center}
{\includegraphics[scale=0.8]{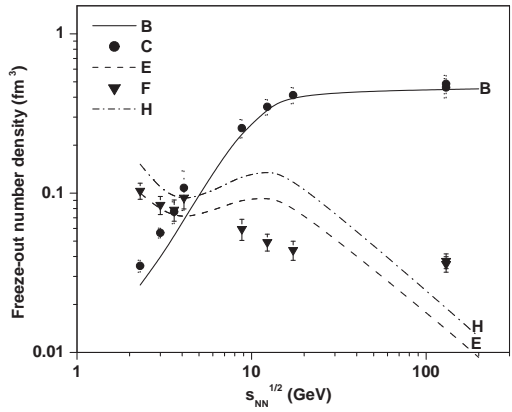}}
\caption{The freeze-out number density is plotted as a function of center-of-mass energy $\sqrt{s_{NN}}$. Curves obtained in the present calculation are compared with the data points obtained by Admova et al., [6]. Solid circular points and triangular points show the HBT data for pion number density and nucleon density at freeze-out, respectively. Solid curve $B$ and dotted curve $E$ show the pion number density and nucleon density at freeze-out predicted by our model, respectively. We have also shown the baryon density in the ideal hadron gas by dash dotted curve $H$ for comparison.}   
\end{center}
\end{figure}
 
\section{RESULTS AND DISCUSSIONS}
     In Fig. 1, we have shown the variation of total freeze-out volume with $\sqrt{s_{NN}}$. Our results amply support the finding that the freeze-out emissions of strange particles take place earlier than the pions. Moreover, the volume first shows a flat minimum around $\sqrt{s_{NN}}\approx5$ GeV and then it rapidly increases. At RHIC energy, the freeze-out volume of the fireball is around $11000$ fm$^3$and it should be far larger than the volume obtained from the radius of gold nucleus used in the collider experiment. In order to compare our results with the HBT data [6], we have also shown in Fig. 1, the fractional volume at chemical freeze-out (corresponding to a slice of one unit of rapidity i.e., $dV/dy$) [19] and its variation with the center-of-mass energy.  This is the volume corresponding to HBT volume. We have calculated $dV/dy$ from the experimental values of $dN/dy$ and dividing it by the theoretical values of number density of the corresponding particle as extracted from the above excluded volume model. As expected, the features of the curves show agreement with each other, we get a minimum in all the curves in agreement with the HBT data. However, the values of $dV/dy$ as obtained in our model are somewhat smaller than the HBT value. This is understandable because the difference arises due to the expansion in the HG between chemical and thermal equilibria. In this connection, we want to emphasize that the definition of freeze-out volume in heavy-ion collisions is a little unclear because the system undergoes a collective expansion which results in a space-momentum correlations of the emitted pions. The observed HBT radii become $k_T$  dependent and the data have been taken at $\langle k_T\rangle$ values around $0.16$ GeV/c. It should be mentioned that the comparison of the freeze-out volume determined from two separate mechanisms should be taken with care because the freeze-out mechanism is dealt in a different fashion in the above pictures.    
   
    Existence of a minimum with respect to the center-of-mass energy in Fig. 1 also warrants an explanation. Such a minimum has been explained by Admova et al., [6] assuming that the matter undergoes a change from baryonically dense medium to mesonically dominant medium. In order to illustrate this, we have also calculated pion density and baryon density at chemical freeze-out and plotted their variations with collider energy in Fig. 2. Our results are in agreement with the data obtained from HBT thermal freeze-out analysis [6]. Fig. 2 indicates that, where the model describes the pion freeze-out densities, it fails to describe the baryon densities at large energies. However, at small $\sqrt{s_{NN}}$, the model fails on the pion densities but describes the baryon densities reasonably well. The large baryon density at RHIC energies still remains an involved problem and our model gives a low density in comparison to the data at these energies. Usually in statistical (thermal model) calculations, one finds that the ideal hadron gas picture (without excluded volume correction) have been used by many authors as a valid description for hadron ratios obtained in heavy ion collisions at RHIC energies because this region is mostly populated by mesons. Therefore, we have also shown in Fig. 2, the baryon density obtained in ideal gas picture. The baryon density curve in ideal gas approximation is large and does not show any agreement with the baryon density data obtained by Admova et al., [6]. This shows that ideal gas picture is unsuitable for the description of the particle densities. Besides we know that a phase transition from hadron gas to QGP does not materialize in the ideal hadron gas picture. Here one must note that the ratio of $dN/dy$ over the HBT volume gives the correct estimate of the freeze-out density. Similarly in the thermal model, total multiplicity $\langle N\rangle$ divided by total volume $V$  gives the freeze-out density. But in this case one assumes a homogeneous freeze-out and particle emissions are homogeneously occurring from the surface of the fireball and the densities are homogeneous throughout the decoupling volume. Attempts have been made to incorporate corrections arising due to inhomogeneties in the homogeneous freeze-out picture. Recently it was shown by Dumitru et al., [28] that no statistical significant improvement over homogeneous freeze-out was observed at SPS and RHIC energies. 
   
    In conclusion, the presence of a large source size and/or a long duration of particle emission resulting from the mixed phase during a first order phase transition from a QGP to a hot and dense hadron gas is indicative of the QGP formation. We find that a chemically equilibrated hadron gas obtained in our model gives a large freeze-out volume of the fireball at RHIC energy and this picture supports the idea of QGP formation before hadronization because a huge size of a homogeneous fireball source can only arise if a mixed phase has occurred before the formation of hot and dense hadronic fireball.  However, one still needs a correct mechanism for incorporating hydrodynamic expansion in both the pictures. An agreement on the features of the results obtained by two completely different mechanisms strengthens our idea regarding the existence of a mixed phase in the ultra-relativistic heavy ion collisions. Moreover, a large baryon density at RHIC energy still poses an anomalous problem which needs an explanation.
  
\section*{ACKNOWLEDGEMENT}

   One of the authors (M. Mishra) would like to acknowledge the financial support as Senior Research Fellowship (SRF) from Council of Scientific and Industrial Research (CSIR), New Delhi, India.

\end{document}